# Entropy Principle in Direct Derivation of Benford's Law

Oded Kafri
Varicom Communications, Tel Aviv 68165 Israel
oded@varicom.co.il

The uneven distribution of digits in numerical data, known as Benford's law, was discovered in 1881. Since then, this law has been shown to be correct in copious numerical data relating to economics, physics and even prime numbers. Although it attracts considerable attention, there is no *a priori* probabilistic criterion when a data set should or should not obey the law. Here a general criterion is suggested, namely that any file of digits in the Shannon limit (namely, having maximum entropy) has a Benford's law distribution of digits.

Bedford's law is an empirical uneven distribution of digits that is found in many random numerical data. Numerical data of natural sources that are expected to be random exhibit an uneven distribution of the first order digits that fits to the equation,

$$\rho(n) = \log_{10}(1+\frac{1}{n}), \text{ where } n = 1,2,3,4,5,6,7,8,9 \quad (1)$$

Namely, digit 1 appears as the first digit at probability $\rho(1)$, which is about 6.5 times higher than the probability $\rho(9)$ of digit 9 (fig 1).

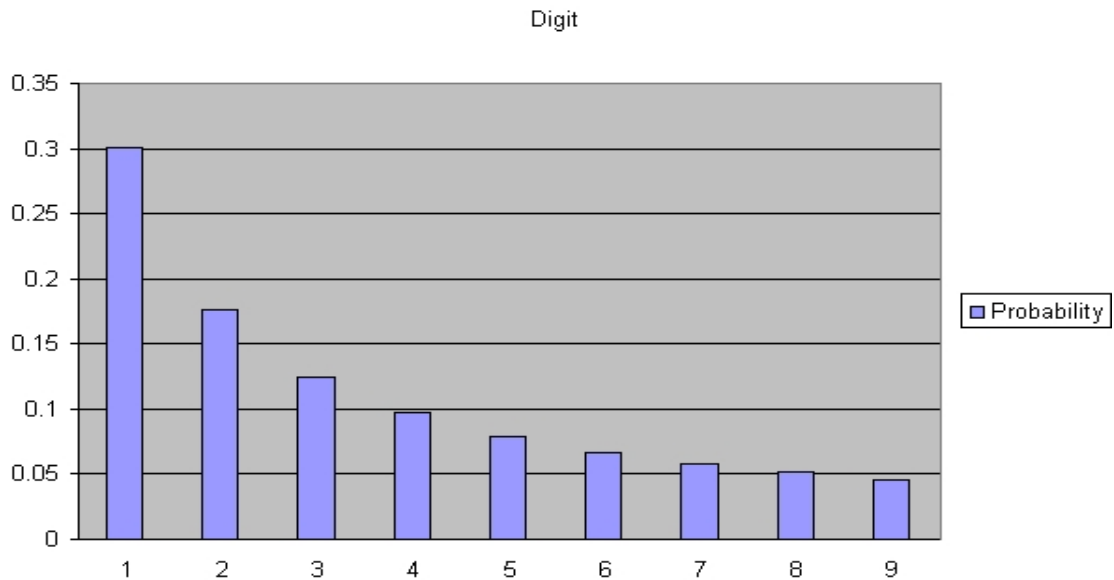

**Fig 1**. *Benford's law predicts a decreasing frequency of first digits, from 1 through 9.*

Eq.(1) was suggested by Newcomb in 1881 from observations of the physical tear and wear of books containing logarithmic tables [1]. Benford further explored the phenomena in 1938, empirically checked it for a wide range of numerical data [2], and unsuccessfully attempted to present a formal proof. Since then Benford's law was found also in prime numbers [3], physical constants, Fibonacci numbers and many more [3,4,5].

Benford's law attracts a considerable attention [6]. Attempts for explanation are based on scale invariance [7] and base invariance [8,9,10] principles. However, there are no *a priori* well-defined probabilistic criteria when a data set should or should not obey the law [4]. Benford's distribution of digits is counterintuitive as one expects that a random numbers would result in uniformity of their digits distribution, namely, $\rho(n) = \frac{1}{9}$ as in the case of an unbiased lottery. This is the reason why Benford's law is used by income tax agencies of several nations and states for fraud detection of large companies and accounting businesses [4,11,12]. Usually, when a fraud is done, the digits are invoked in equal probabilities and the distribution of digits does not follow Eq. (1).

In this paper Benford's law is derived according to a standard probabilistic argumentation. It is assumed that, counter to common intuition (that digits are the logical units that comprise numbers) that the logical units are the 1's. For example, the digit 8 comprises of 8 units 1 etc. This model can be easily viewed as a model of balls and boxes, namely:

A) Digit *n* is equivalent to a "box" containing *n* none-interacting balls.

B) *N* sequence of such "boxes" is equivalent to a number or a numerical file.

C) All possible configurations of the boxes and balls, for a given number of balls, have equal probability.

The last assumption is the definition of equilibrium and randomness in statistical physics. In information theory it means that the file is in the Shannon limit (a compressed file).

A number is written as a combination of ordered digits assuming a given base *B*. When we have a number with *N* digits of base *B*, we can describe the number as a set of $N$ boxes, each contains a number of balls $n$, when $n$ can be any integer from 0 to $B-1$. We designate the total number of balls in a number as $P$.

An unbiased distribution of balls in boxes means an equal probability for any ball to be in any box. Hereafter, it is shown that this assumption is equivalent to assumption C and yields Benford law.

The "intuitive" distribution in which each box has an equal probability to have any digit $n$ ($n$ balls) does not means an equal probability for any single ball to be in any box, but an equal probability for any group of $n$ balls (the digit $n$) to be in any box.

For example, for base $B$=4 there are four digits 0,1,2,3. The highest value of a 3 digits number in this base is 3|3|3, which contains 9 balls. There is only one possible configuration to distribute the 9 balls in 3 boxes (because the limit of 3 balls per box). However, in the case of 3 balls in 3 boxes there are several possible configurations, namely: 3|0|0, 0|3|0, 0|0|3, 2|1|0, 2|0|1, 1|2|0, 0|2|1, 1|0|2, 0|1|2, and 1|1|1. We see that digit 1 appears 9 times, digit 2 appears 6 times, and digit 3 appears 3 times. It is worth noting that the ratio of the digits 9:6:3, $\rho(1)=0.5$, $\rho(2)=0.33\bar{3}$, and $\rho(3)=0.16\bar{6}$, is independent of $N$, which is the reason why 0 is not included in Benford's law. As we see in the example above, each box has the same probability of having 1, 2 or 3 balls as the other boxes, however, the probability of a box of having 3 balls is smaller than the probability of a box to have 2 balls and the probability of a box of having one ball is the highest. The reason for this is that in order for a box to have several balls it has to score a ball several times, since the probability of a box to score $n$ balls is smaller as $n$ increases, lower value digits have higher probability. The formal calculation of the distribution of balls in the boxes was done by writing all possible ten configurations (in general

$\dfrac{(N+P-1)!}{P!(N-1)!}$ configurations), and give each one of them an equal probability) and then counting the total number of each digit, regardless of its location.

The distribution of $P$ balls in $N$ boxes in equilibrium is a classic thermodynamic problem. Equilibrium is defined in statistical mechanics as a statistical ensemble in which all the possible configurations have an equal probability. The equilibrium distribution function $\rho(n)$ (the fraction of boxes having $n$ balls) is calculated in a way that it yields maximum entropy, which means equal probability for all the configurations (microstates).

The standard way to find the distribution function in equilibrium is to maximize entropy $S$ ($S$ is proportional to the logarithm of the number of configurations $\Omega$), under the constraint of a fixed number of balls $P$.

To do this, we apply the Stirling approximation; since $\Omega(N,P) = \dfrac{(N+P-1)!}{(N-1)!P!}$ we obtain,

$$S = \ln \Omega \cong N\{(1+n)\ln(1+n) - n\ln n\} \qquad (2)$$

Where $n = \dfrac{P}{N}$. The number $n$ is any integer (limited by $B-1$). If we designate the number of boxes having $n$ balls by $\phi(n)$ then $P = \sum_{n=1}^{B-1} n\phi(n)$.

It should be noted that we count all the boxes in all the configurations excluding the empty boxes (the 0's) and the boxes that contain more balls then $B-1$.

To find $\phi(n)$ that maximizes $S$ we will use the Lagrange multipliers method namely to define a function $f(n) = N\{(1+n)\ln(1+n) - n\ln n\} - \beta(P - \sum_{n=1}^{B-1} n\phi(n))$. The first term is Shannon entropy and the second term is the conservation of the balls. $\beta$ is the Lagrange multiplier. To find $\phi(n)$ we substitute $\dfrac{\partial f(n)}{\partial n} = 0$, hence

$$\phi(n) = \frac{1}{\beta}\ln(\frac{n+1}{n}). \qquad (3)$$

We are interested in the normalized distribution, namely,

$$\rho(n) = \frac{\phi(n)}{\sum_{n=1}^{B-1}\phi(n)} \qquad (4)$$

Since $\sum_{n=1}^{B-1}\phi(n) = \dfrac{1}{\beta}\ln B$ it follows that

$$\rho(n) = \frac{\ln(1+\frac{1}{n})}{\ln B} = \log_B(1+\frac{1}{n}) \qquad (5)$$

namely, Benford's law.

In the normalization of $\phi(n)$ the quantity $\dfrac{N}{\beta}$ disappeared. That means that the distribution function is independent in the digit location in *P* and *N* and it is only a function of *B*. That is the reason why Benford law is so general.

Reconsidering the example of 3 balls in 3 boxes, we calculate from Eq,(5) that $\rho(1) = 0.5$ ; $\rho(2) \cong 0.29$ $\rho(3) \cong 0.21$. The total number of the none-zero digits (1, 2 and 3) is 18, and the distribution points to the ratio 9:5:4 as compared to the result of 9:6:3 that was obtained in the numerical example. The deviation from the theoretical calculation is explained by the fact that Sterling approximation yields a better fit as the number of digits grows.

Benford's law distribution was shown recently to be a special case of Planck distribution of photons at a given frequency [13]. It is intriguing that digits distribution of prime numbers also obeys the Planck statistics.

**Acknowledgment**: This paper was written due to the encouragement of Alex Ely Kossovsky who is a truly a Benford's law expert and enthusiast. I acknowledge his most valuable comments.